\documentstyle[a4,12pt,epsf]{article}

\begin{document}

\newcommand{\Lslash}[1]{ \parbox[b]{1em}{$#1$} \hspace{-0.8em}
                         \parbox[b]{0.8em}{ \raisebox{0.2ex}{$/$} }    }
\newcommand{\Sslash}[1]{ \parbox[b]{0.6em}{$#1$} \hspace{-0.55em}
                         \parbox[b]{0.55em}{ \raisebox{-0.2ex}{$/$} }    }
\newcommand{\Mbf}[1]{ \parbox[b]{1em}{\boldmath $#1$} }
\newcommand{\mbf}[1]{ \parbox[b]{0.6em}{\boldmath $#1$} }
\newcommand{\beq}{\begin{equation}}
\newcommand{\eeq}{\end{equation}}
\newcommand{\beqa}{\begin{eqnarray}}
\newcommand{\eeqa}{\end{eqnarray}}
\newcommand{\skipfields}{\!\!\!\!\! & \!\!\!\!\! &}
\newcommand{\half}{\frac{1}{2}}
\newcommand{\gsim}{\buildrel > \over {_\sim}}
\newcommand{\lsim}{\buildrel < \over {_\sim}}
\newcommand{\ie}{{\it ie}}
\newcommand{\eg}{{\it eg}}
\newcommand{\cf}{{\it cf}}
\newcommand{\bfp}{{\bf p}}
\newcommand{\bfq}{{\bf q}}
\newcommand{\bfk}{{\bf K}}
\newcommand{\bfklc}{{\bf k}}
\newcommand{\bfb}{{\bf b}}
\newcommand{\bfj}{{\bf j}}
\newcommand{\bfr}{{\bf r}}
\newcommand{\bfR}{{\bf R}}
\newcommand{\bfeps}{{\mbox{\boldmath $\varepsilon$}}}
\newcommand{\bfgam}{{\mbox{\boldmath $\gamma$}}}
\newcommand{\etal}{{\it et al.}}
\newcommand{\gev}{{\rm GeV}}
\newcommand{\jpsi}{J/\psi}
\newcommand{\order}[1]{{\cal O}(#1)}
\newcommand{\eq}[1]{Eq.\ (\ref{#1})}
\newcommand{\ptr}{p_T}
\newcommand{\as}{\alpha_s}
\newcommand {\pom}{\rm {I\hspace{-0.2em}P}}
\newcommand {\xpom} {\mbox{$x_{_{\pom}}$}}

\newcommand{\PL}[3]{Phys.\ Lett.\ {\bf {#1}}, {#2} ({#3})}
\newcommand{\NP}[3]{Nucl.\ Phys.\ {\bf {#1}}, {#2} ({#3})}
\newcommand{\PRD}[3]{Phys.\ Rev.\ D {\bf {#1}}, {#2} ({#3})}
\newcommand{\PR}[3]{Phys.\ Rev.\  {\bf {#1}}, {#2} ({#3})}
\newcommand{\PRL}[3]{Phys.\ Rev.\ Lett.\ {\bf {#1}}, {#2} ({#3})}
\newcommand{\ZPC}[3]{Z. Phys.\ {\bf C{#1}}, {#2} ({#3})}
\newcommand{\PRe}[3]{Phys.\ Rep.\ {\bf {#1}}, {#2} ({#3})}

\begin{titlepage}
\begin{flushright}
        NORDITA--96/68 P \\
        SLAC--PUB--7342 \\
        hep-ph/9611278 \\
       \today
\end{flushright}

\vskip .5cm

\centerline{\Large \bf Rapidity gaps in perturbative QCD}

\vskip .5cm
\centerline{\bf Stanley J. Brodsky\footnote{
Work supported in part by Department of Energy contract 
DE--AC03--76SF00515 and DE--AC02--76ER03069. }}
\centerline{\sl Stanford Linear Accelerator}
\centerline{\sl Stanford University, Stanford, California 94309, USA}

\vskip .5cm

\centerline{\bf Paul Hoyer\footnote{Work supported in part by the EU/TMR
contract ERB FMRX--CT96--0008.} and Lorenzo Magnea\footnote{On leave from
Universit\`a di Torino, Italy}}
\centerline{\sl NORDITA}
\centerline{\sl Blegdamsvej 17, DK--2100 Copenhagen \O, Denmark}

\vskip 1cm

\begin{abstract}

We analyze diffractive deep inelastic
scattering within perturbative QCD by studying lepton scattering
on a heavy quark target.
Simple explicit expressions are derived in impact parameter space for
the photon wave function and the scattering cross sections corresponding to
single and double Coulomb gluon exchange.
At limited momentum transfers to the target, the results agree with
the general features of the ``aligned jet model''.
The color--singlet exchange cross section
receives a leading twist contribution only from the
aligned jet region, where the transverse size of the photon wave function
remains finite in the Bjorken scaling limit. In contrast to inclusive DIS,
in diffractive events there is no leading twist contribution to
$\sigma_L/\sigma_T$ from the lowest order $(q\bar q)$ photon Fock state, and
the cross section for heavy quarks is power suppressed in the quark mass.
There are also important contributions with large momentum transfer to
the target, which corresponds to events having high transverse momentum
production in both the projectile and target rapidity regions, separated
by a rapidity gap.
\end{abstract}

\end{titlepage}

\newpage
\renewcommand{\thefootnote}{\arabic{footnote}}
\setcounter{footnote}{0}
\setcounter{page}{1}

\section{Introduction}
\label{intro}

The physics of rapidity gaps in deep inelastic lepton scattering (DIS) has
been qualitatively described using the ``aligned jet'' model
\cite{aljet}. This model forms the basis for several QCD studies of
diffractive DIS, where various assumptions have been made concerning the
interaction with the target (Pomeron exchange~\cite{pom}, BFKL
ladders~\cite{bfkl}, soft color fields~\cite{class}).
In this paper we wish to test and clarify the physics of the aligned jet
model by choosing a particularly simple target, which allows a full
perturbative  calculation. We believe that such an explicit (and
straightforward) QCD  calculation can serve to illustrate several points of
principle, and help  distinguish between the physics of the photon projectile
and the target  ``diffractive structure function''.

In the aligned jet model, the virtual photon breaks up into an asymmetric
quark--antiquark pair. In the frame where the target is at rest, one of the
two partons, say the quark, takes nearly all the photon energy $\nu$, and
forms the ``current jet''. The antiquark has then a finite energy of
$\order{\Lambda_{QCD}/x}$, even in the scaling limit, $Q^2,\ \nu \to
\infty$ with $x=x_{Bj}=Q^2/2m_N\nu$ fixed. The antiquark also has low
transverse momentum $p_\perp = \order{\Lambda_{QCD}}$ with respect to the
photon direction, and scatters softly on the target. (Conversely, in the frame
in which the target has infinite momentum, the antiquark is seen as a
quark emerging from the target with momentum fraction $x$, and the soft
scattering probability corresponds to the structure function.)

Since the antiquark momentum is proportional to $1/x$, it reaches quite high
values in the small $x$ range accessible at Hera. Thus, in analogy to hadron
scattering, also the antiquark scattering is expected to develop a diffractive
(color--singlet exchange) part, resulting in a ``rapidity gap'' between the
antiquark and target.

The soft target scattering cannot be calculated in perturbation theory. In
inclusive DIS, one relies on the QCD factorization theorem to separate the
hard and soft momenta in the process. For diffractive DIS, however, there
is no analogous proof of the separation of hard and soft subprocesses, so
that the analysis is more model dependent and may provide new insights
into the structure of hadrons and nuclei (see for example Ref.~\cite{frstr}).

Here we wish to study a very simple model for the target scattering by
assuming that the target is a heavy quark~\cite{dbh}.
This selects Coulomb gluon exchange as the dominant high energy interaction.
We will consider both one and two--gluon exchange
contributions in order to model the inclusive and
diffractive DIS, respectively. Our approach can be viewed in two ways:

\begin{itemize}
\item[(i)] As a toy model, consistent with PQCD (in fact it is PQCD, as the
process we consider is theoretically conceivable). Many of our results will
be similar to calculations involving more sophisticated assumptions about
the target scattering, for example in terms of BFKL gluon
ladders~\cite{bfkl}.
We feel that it is useful nevertheless to explicitly demonstrate those
features which follow just from the photon structure through the aligned
jet mechanism, by making the simplest possible assumption about the target
scattering.

\item[(ii)] As the lowest order of a more complete calculation which also
takes into account gluon radiation and scattering. As noted above, in the
target rest frame the antiquark is seen as part of the photon wave function;
similarly, the (high energy part of the) gluon ladder to the target is built
from gluons radiated from the initial quark pair.
It is thus not unreasonable to expect the nature (and mass) of the target
itself to be rather immaterial once sufficiently many gluons are considered
in the photon wave function.
\end{itemize}

At small $x$ and, consequently, large longitudinal momenta $p_z
\sim\order{\Lambda_{QCD}/x}$, the parton interaction times in
the target (which are of $\order{1\ {\rm fm}}$) are short compared to
their (Lorentz--dilated) lifetimes. Hence the scattering
amplitude factorizes into a Fock state formation amplitude (the photon wave
function) and a scattering amplitude for each
Fock state~\cite{nz,kopo,BHQ}.
Moreover, the scatterer can move only a negligible transverse
distance $\sim Lp_\perp/p_z$ in a soft scattering on a target of
length $L$. Hence the scattering
amplitude is approximately diagonal in impact parameter space. We show
explicitly how these features arise in the PQCD amplitude for one and two
gluon exchange at leading order in $1/x$. These features, which follow from
general quantum mechanical principles, must be present to arbitrary orders in
perturbation theory, providing an important conceptual and technical
simplification of higher order calculations.

Since hadronization is an inherently non--perturbative phenomenon, it is not
possible to prove that the rapidity gap between the target and diffractive
scatterer is maintained in the evolution to the hadronic final state.
Nevertheless, one can take a step in this direction by calculating the
distribution of soft gluon radiation. This is analogous to the PQCD ``proof''
of the string effect in $e^+e^- \to q\bar q g \to$
hadrons~\cite{string}. From the structure of the photon
wave function we verify that soft gluon radiation is indeed dominantly
emitted in the rapidity region between the quark and the antiquark.
We expect this feature to survive a scattering with color--singlet exchange,
leaving a rapidity gap between hadrons in the photon and target
fragmentation regions.

An immediate consequence of our analysis is that the dependence
of the single--gluon exchange cross section on the photon energy $\nu$
dominates the multi--gluon exchange contribution by a factor of $\order{\log
\nu}$, a feature that appears also in QED~\cite{bethemax}. This
logarithm is due to scattering at large impact parameters, which for
a color--singlet target will saturate at impact parameters of the order
of the target radius.

Our model also shows that there are important contributions to hard
diffractive scattering from small impact parameters of $\order{1/Q}$.
This should show up in $ep$ scattering as events where both the projectile
and the target fragment into large transverse momentum particles, with a
rapidity gap between the projectile and target fragments.

The purpose of this paper is to present a self--contained perturbative model
of diffractive DIS
which bypasses the uncertainties due to Pomeron physics.
With this in mind, we shall be rather explicit and assume no methods beyond
the standard arsenal of perturbative QCD.

\section{Scattering amplitudes on a massive target}
\label{ampsec}

In this section we will state the kinematic properties of our model, and
present the calculation of the relevant amplitudes. We shall be concerned
with the QCD amplitude for
\beq
\gamma^*(q) + t(m_t) \rightarrow  q(p_1) + \bar{q}(p_2) + t(m_t)~~,
\label{process}
\eeq
where we take the target $t$ to be a heavy quark of mass $m_t$. We shall
calculate the process (\ref{process}) in the small--$x$ limit\footnote{We
use the standard definition of $x=Q^2/2m_N\nu$, $m_N$ being the nucleon mass.
All our results will depend only on the combination $xm_N=Q^2/2\nu$ and are
insensitive to the value of the (large) target mass $m_t$.},
at leading order in PQCD, and study both inclusive and diffractive
DIS, the latter being defined by the requirement that the $q
\bar{q}$ pair be in a color--singlet state after the scattering. In both cases,
we demonstrate how the amplitude factorizes into the photon wave function (as
derived in the Appendix) and the $q\bar q$ scattering amplitude.

The calculation is greatly simplified by
assuming a large target mass $m_t$,
which implies Coulomb gluon exchange and negligible energy transfer. The
large target mass limit is thus a natural laboratory for perturbative studies
of DIS and for the emergence of rapidity gaps. Since, as we shall see, much of
the physics is encoded in the photon wave function, there is reason  to
hope that many features of the process will be target--independent  and
generally
applicable to the physical case of a color--singlet, finite mass target.

\begin{figure}[htbp]
\begin{center}
\leavevmode
{\epsfxsize=10truecm \epsfbox{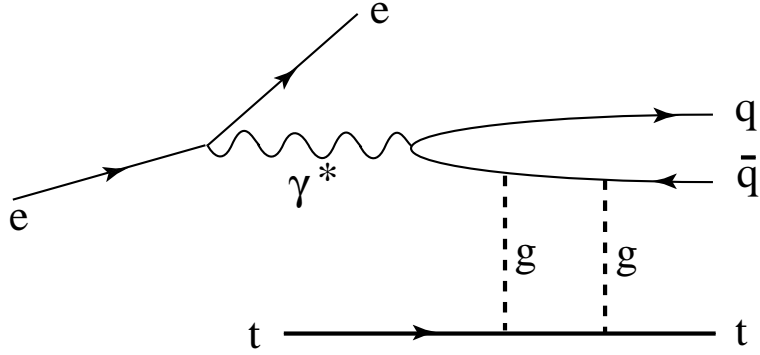}}
\end{center}
\caption[*]{A diagram contributing to the process
$e + t(m_t) \rightarrow  q(p_1) + \bar{q}(p_2) + t(m_t)$ with double--gluon
exchange.}
\label{fig1}
\end{figure}

The space--time picture we have in mind, depicted in Fig. 1 for double--gluon
exchange, is thus the  following. A high energy photon of energy
$\nu$ and virtuality $Q^2$ fluctuates into a $q \bar{q}$ pair a long time
($\order{\nu/Q^2}$) before reaching the target. The pair subsequently
interacts  with the massive target via single or multiple Coulomb gluon
exchange, with a limited total momentum transfer ${\bf K}$. For a diffractive
process we require the pair to emerge from the scattering in a color--singlet
state, so that a rapidity gap between the pair and the target can be formed.

We will work in the target rest frame in which the virtual photon is moving
along the positive $z$ axis, $q = (\nu,{\bf 0}_\perp,\sqrt{\nu^2+Q^2})$. The
small value of Bjorken $x$ implies that the photon is off--shell by an amount
which is  small compared with its energy,
\beq
q_z \simeq \nu + Q^2/2\nu~~.
\label{nu}
\eeq
We take all transverse momenta to be much smaller than the
longitudinal ones. The 3--momenta of the quarks are parametrized as
\beqa
\bfp_1 & = & (\bfp_{1\perp},z q_z)~~,  \nonumber \\
\bfp_2 & = & (\bfp_{2\perp},(1-z) q_z + K_z)~~,
\label{pairmom}
\eeqa
where ${\bf p}_{2\perp} = {\bf K}_\perp - {\bf p}_{1\perp}$.
Energy transfer to the target is suppressed by powers of the target mass, so
that energy conservation reads
\beq
\nu \simeq q_z + K_z + \frac{\bfp^2_{1\perp} + m^2}{2 z \nu}
      + \frac{\bfp^2_{2\perp} + m^2}{2 (1 - z) \nu}~~.
\label{encons}
\eeq
This implies that the total longitudinal momentum $K_z$ transferred to the
target is small, of order $Q^2/\nu$ or $p^2_\perp/\nu$.

With these kinematic preliminaries in mind, we turn to the evaluation of
the scattering amplitudes. In particular, using the standard decomposition
of the total amplitude given in the Appendix, \eq{apol}, we concentrate on
the amplitude for  process (\ref{process}) with a virtual photon of definite
polarization,
\beq
A(\lambda;\lambda_1,\lambda_2) = \bfeps(\lambda) \cdot
\bfj_h (\lambda_1,\lambda_2)~~,
\label{comptamp}
\eeq
where $\lambda_1$ and $\lambda_2$ are the helicities of the quark and the
antiquark, respectively.

We first calculate the single Coulomb gluon exchange amplitude $A_1$,
which in many respects is analogous to the two--gluon (color--singlet
exchange) amplitude $A_2$. In particular, both amplitudes will turn out to
factorize in impact parameter space into a product of the photon wave function
(\cf\ Appendix) and the
$q\bar q$ scattering amplitude.

\subsection{One--gluon exchange}
\label{oge}

The scattering amplitude is the sum of the two diagrams given in Fig. 2.
Consider for example the first diagram, in which the Coulomb gluon is
attached to the antiquark. In the limit of large mass $m_t$ the target
contributes only a factor $2m_t$ times the color factor $(T^a)_{CD}$.
The fermion propagator in the upper vertex may be split into forward and
backward propagating parts using the identity
\beq
\frac{\Sslash{p}+m}{p^2-m^2+i\epsilon}=\frac{1}{2E}\left[\frac{E\gamma^0
-\bfp\cdot\bfgam+m}{p^0-E+i\epsilon}+\frac{E\gamma^0+\bfp\cdot\bfgam-m}
{p^0+E-i\epsilon}\right]~~, \label{fident}
\eeq
where $E=\sqrt{\bfp+m^2}$. This procedure splits the covariant
diagram into different time orderings of the vertices, reproducing the
results of time--ordered
and light--cone perturbation theory.
In the DIS limit, the fermion
energies are $\gsim\order{(\bfp_\perp^2+m^2)2\nu/Q^2}$, hence for small $x$
we may neglect the backward scattering term in \eq{fident}, and the
denominator simplifies to
\beq
2 E (p_0 - E) \simeq
- \frac{\bfp_{1\perp}^2+\varepsilon^2}{z} \label{enden}
\eeq
where
\beq
\varepsilon^2 = m^2 + z (1 - z) Q^2~~. \label{eps}
\eeq
The amplitude for Fig. 2a is then
\beqa
A_1^{(1)}(\bfp_{1\perp},\bfk_\perp) & = &
e e_q g^2~(T^a)_{AB}~(T_a)_{CD}~\frac{2 m_t z}{\bfk^2 (\bfp^2_{1\perp}
+ \varepsilon^2)} \label{toamp1} \\
& \times & \sum_\beta \left[ \bar{u}_{\lambda_1} (p_1)
\bfeps(\lambda) \cdot \bfgam v_\beta(p_2 - K) ~\bar{v}_\beta(p_2 - K)
\gamma_0 v_{\lambda_2}(p_2) \right]~~. \nonumber
\eeqa
Using the explicit expressions for the spinors given in the Appendix, we
have in the high energy limit
\beq
\bar{v}_\beta(p_2 - K) \gamma_0 v_{\lambda_2}(p_2) =
2 (1 - z) \nu \delta_{\beta,\lambda_2}~~.
\label{collspin}
\eeq
The second diagram is treated similarly, giving
\beqa
A_1^{(1)}(\bfp_{1\perp},\bfk_\perp) & = &
e e_q g^2~(T^a)_{AB}~(T_a)_{CD}~\frac{4 m_t \nu z (1 - z)}{\bfk^2
(\bfp^2_{1\perp} + \varepsilon^2)} \nonumber \\
& \times & \bar{u}_{\lambda_1} (p_1) \bfeps(\lambda) \cdot \bfgam
v_{\lambda_2}(p_2 - K)~~, \label{1gamps} \\
A_1^{(2)}(\bfp_{1\perp},\bfk_\perp) & = &
- ~e e_q g^2~(T^a)_{AB}~(T_a)_{CD}~\frac{4 m_t \nu z (1 - z)}{\bfk^2
(\bfp^2_{2\perp} + \varepsilon^2)} \nonumber \\
& \times & \bar{u}_{\lambda_1} (p_1 - K) \bfeps(\lambda) \cdot \bfgam
v_{\lambda_2}(p_2)~~. \nonumber
\eeqa

\bigskip
\begin{figure}[htbp]
\begin{center}
\leavevmode
{\epsfxsize=10truecm \epsfbox{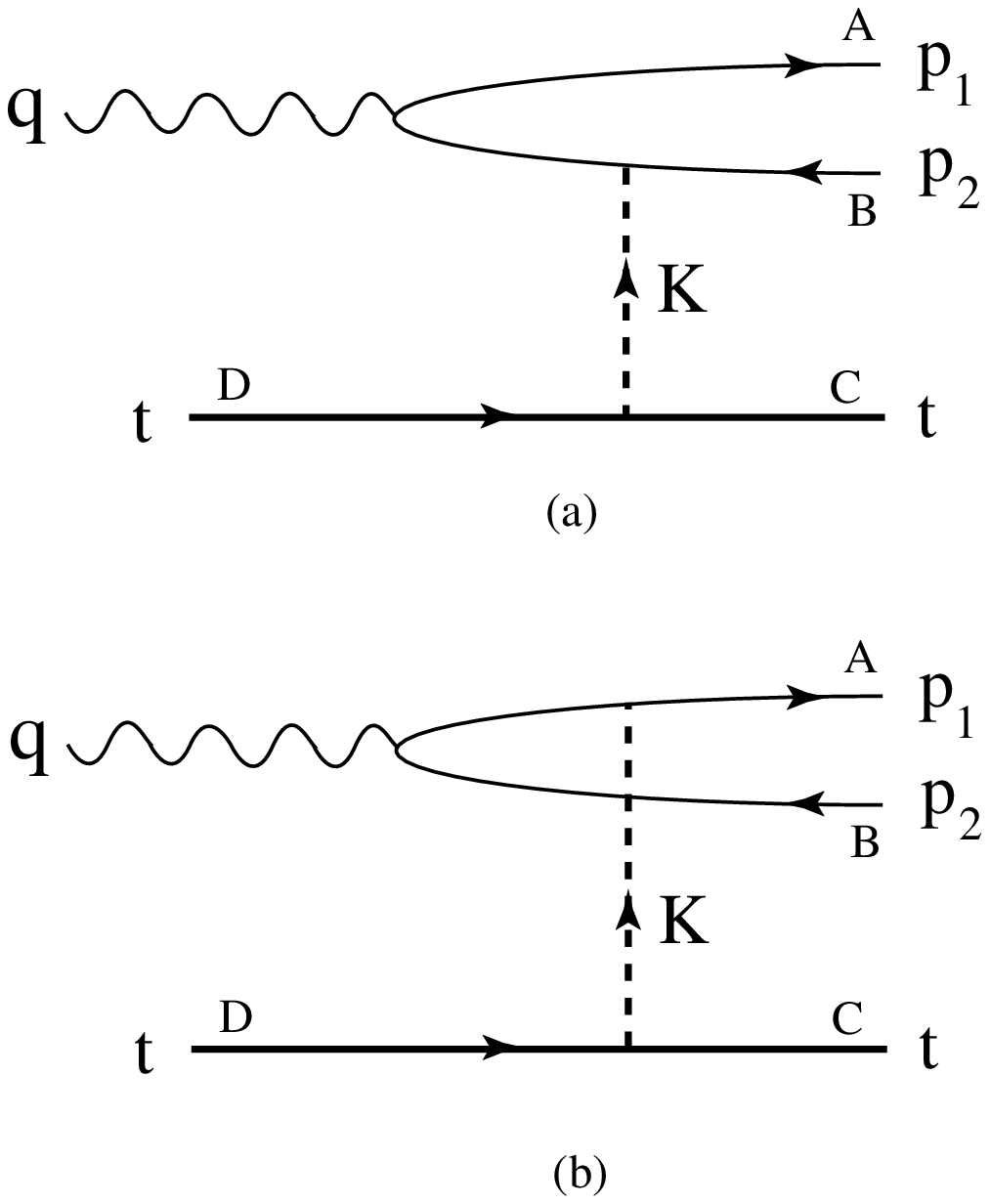}}
\end{center}
\caption[*]{Single--gluon exchange diagrams contributing to
inclusive deep inelastic scattering on a heavy quark target.}
\label{fig2}
\end{figure}

The close relationship between these amplitudes and the photon wave
function given in the Appendix, \eq{awfscal}, is already apparent.
Fourier transforming to transverse distance space $\bfr_\perp$ between the
quarks, by defining
\beq
{A}(\bfr_\perp,\bfk_\perp) = \int
\frac{d^2 \bfp_{1\perp}}{(2 \pi)^2} A(\bfp_\perp,\bfk_\perp)
\exp({\rm i} \bfr_\perp \cdot \bfp_{1\perp})~~,
\label{toimpa}
\eeq
we find consequently that the one--gluon exchange amplitude
\beq
{A}_1(\bfr_\perp,\bfk_\perp) = {\rm i} g^2 2m_t (T^a)_{AB}~(T_a)_{CD}
W_1(\bfr_\perp,\bfk_\perp)
V^{\lambda_1,\lambda_2}_\lambda (z, \bfr_\perp)~~,
\label{simprel1}
\eeq
is proportional to the photon wave function $V^{\lambda_1,
\lambda_2}_\lambda (z, \bfr_\perp)$ of \eq{bampl}. The eikonal
factor
\beq
W_1(\bfr_\perp,\bfk_\perp) = \frac{1 - \exp({\rm i}
\bfr_\perp \cdot \bfk_\perp)}{\bfk_\perp^2}
\label{eik1}
\eeq
is independent of the photon helicity $\lambda$ and of the quark helicities
$\lambda_1,\lambda_2$. Since all dependence on the momentum transfer
$\bfk_\perp$ to the target is in $W_1$, we may make a further transformation
to target impact parameter ($\bfR_\perp$) space by defining
\beqa
{W}_1(\bfr_\perp,\bfR_\perp) & = &
\int \frac{d^2 \bfk_{\perp}}{(2 \pi)^2} W_1(\bfr_\perp,\bfk_\perp)
\exp({\rm i} \bfR_\perp \cdot \bfk_{\perp}) \nonumber \\
& = & \frac{1}{2\pi} \lim_{\epsilon\to 0} \left[K_0(|\bfR_\perp|\epsilon) -
K_0(|\bfR_\perp+\bfr_\perp|\epsilon)\right] \nonumber \\
& = & \frac{1}{2\pi} \log\left(\frac{|\bfR_\perp+\bfr_\perp|}{|\bfR_\perp|}
\right) \label{besseleik}
\eeqa

The physical interpretation of our result, \eq{simprel1}, is straightforward.
The $q \bar{q}$ pair, which is in a color--singlet state before the
scattering, forms a color dipole of size $\bfr_\perp$. This dipole cannot be
detected by a gluon with $|\bfk_\perp| \lsim 1/|\bfr_\perp|$, in which case
the two diagrams tend to cancel. Thus in \eq{besseleik}, $W_1 \propto
|\bfr_\perp|/|\bfR_\perp|$ when this ratio is small.
In particular, the infinite Coulomb phases associated with the quark and the
antiquark cancel, and the eikonal factor is their finite remainder. In QED
this factor would simply exponentiate upon adding extra virtual Coulomb
photons. In non--abelian QCD the situation is complicated by color dynamics.
The simple structure of \eq{simprel1} will, however, be preserved for
more than one Coulomb gluon, in the color--singlet exchange channel.

\subsection{Two--gluon exchange}
\label{tge}

Four diagrams contribute to the scattering
amplitude via two--gluon exchange, corresponding to different attachments of
the gluons to the quark and antiquark. The individual diagrams
are infrared divergent, but their sum is finite because of the
dipole cancellation discussed above. We regulate the
infrared divergences by allowing only a finite time $\tau$ to elapse between
the two  gluon exchanges. With a more realistic model for the target, $\tau$
would play the role of  the target size, and we would expect $\tau \sim
O(1/\Lambda_{QCD})$. Note that the true time interval between the two gluon
exchanges is shorter than either the lifetime of the $q\bar q$
fluctuation or its subsequent hadronization time \cite{nz,lmrt}. Our
infrared cutoff $\tau$ implies, as one can see by translating to
time--ordered perturbation theory,  an extra factor
$1 - \exp (- {\rm i} K_{1z} \tau)$ in the amplitudes, where $K_{1z}$ is
the longitudinal momentum of one of the exchanged gluons. As we will see, the
results are not sensitive to the specific value of the cutoff, which can be
removed from the sum of the four diagrams.

Let us consider the diagram in Fig. 3, where the two gluons
with momenta $\bfk_1$ and $\bfk_2$ ($\bfk_1 + \bfk_2 = \bfk$) interact
with the antiquark line. Upon using the identity (\ref{fident}) on the
fermion line between the gluon exchanges we encounter the energy denominator
\beq
\Delta E = - K_{1z} - \frac{1}{2 z (1 - z)\nu} \left[ \varepsilon^2 +
\bfp^2_{1\perp} + z (\bfk^2_{1\perp} -
2 \bfp_{1\perp} \cdot \bfk_{1\perp} ) \right]~~.
\label{endon2}
\eeq
Since $K_{1z}\gsim \order{1/\tau}$ due to our cut--off, we have
$\Delta E \simeq - K_{1z}$ at high energy $\nu$.

\bigskip
\begin{figure}[htbp]
\begin{center}
\leavevmode
{\epsfxsize=10truecm \epsfbox{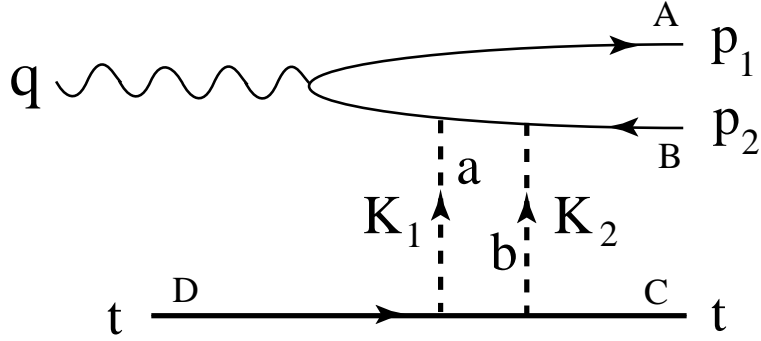}}
\end{center}
\caption[*]{A two--gluon exchange diagram contributing to
diffractive deep inelastic scattering on a heavy quark target.}
\label{fig3}
\end{figure}

It is legitimate to take the target mass limit $m_t \to \infty$
inside the loop integral, since the integral is ultraviolet convergent.
Then the target contributes just an overall factor of $2 m_t$ and a color
coefficient, as for one--gluon exchange. The upper vertex also simplifies
in the high energy limit, and finally we find
\beqa
A_2^{(1,1)} (\bfp_{1\perp},\bfk_\perp) & = &
e e_q g^4~(T^a T^b)_{AB}~(T_b T_a)_{CD}~\frac{4 m_t\nu z (1 -
z)}{\bfp^2_{1\perp} + \varepsilon^2}
\label{a11} \\
& \times & \bar{u}_{\lambda_1} (p_1) \bfeps(\lambda) \cdot \bfgam
v_{\lambda_2} (q - p_1) \int \frac{d^2 \bfk_{1\perp}}{(2 \pi)^2}
J (K_{1\perp},K_{2\perp},\tau)~~,
\nonumber
\eeqa
where $J$ is the integral over the longitudinal component of the loop
momentum. Using the fact that $K_z = K_{1z} + K_{2z}$ is negligible, the
longitudinal momentum integral can be computed exactly,
\beqa
J(a,b,\tau) & = & \int_{- \infty}^{+ \infty} \frac{d K_{1z}}{2 \pi}
\frac{1 - {\rm e}^{- {\rm i} K_{1z} \tau}}{(K_{1z}^2 + a^2)(K_{1z}^2 + b^2)(
K_{1z} - i\epsilon)}
\nonumber \\
& = & \frac{{\rm i}}{2}~\frac{1}{b^2 - a^2} \left[
\frac{1 - {\rm e}^{ - a \tau}}{a^2} - \frac{1 - {\rm e}^{ - b \tau}}{b^2}
\right]~~.
\label{Jint}
\eeqa
Notice that for large values of $\tau$ this becomes simply
\beq
J(a,b,\tau) \sim \frac{{\rm i}}{2} \frac{1}{a^2 b^2}~~,
\label{simpJ}
\eeq
independent of $\tau$.

The other three diagrams can be treated in the same way. The longitudinal
momentum dependence is encoded in the same integral, \eq{Jint}, for
all of them. We find
\beqa
A_2^{(1,2)} (\bfp_{1\perp},\bfk_\perp) & = &
- ~e e_q g^4~(T^b T^a)_{AB}~(T_b T_a)_{CD} \int \frac{d^2
\bfk_{1\perp}}{(2 \pi)^2}
\frac{4 m_t\nu z (1 - z)}{(\bfp_{1\perp} - \bfk_{2\perp})^2 +
\varepsilon^2}  \nonumber \\
& \times & \bar{u}_{\lambda_1} (p_1 - K_2) \bfeps(\lambda) \cdot \bfgam
v_{\lambda_2} (p_2 - K_1)
J (K_{1\perp},K_{2\perp},\tau)~~, \nonumber \\
A_2^{(2,1)} (\bfp_{1\perp},\bfk_\perp) & = &
- ~e e_q g^4~(T^a T^b)_{AB}~(T_b T_a)_{CD} \int \frac{d^2
\bfk_{1\perp}}{(2 \pi)^2}
\frac{4 m_t\nu z (1 - z)}{(\bfp_{1\perp} - \bfk_{1\perp})^2 + \varepsilon^2}
\nonumber \\
& \times & \bar{u}_{\lambda_1} (p_1 - K_1) \bfeps(\lambda) \cdot \bfgam
v_{\lambda_2} (p_2 - K_2)
J (K_{1\perp},K_{2\perp},\tau)~~,  \nonumber \\
A_2^{(2,2)} (\bfp_{1\perp},\bfk_\perp) & = &
e e_q g^4~(T^b T^a)_{AB}~(T_b T_a)_{CD} \frac{4 m_t\nu z (1 -
z)}{\bfp^2_{2\perp} + \varepsilon^2}
\nonumber \\
& \times & \bar{u}_{\lambda_1} (q - p_2) \bfeps(\lambda) \cdot \bfgam
v_{\lambda_2} (p_2) \int \frac{d^2 \bfk_{1\perp}}{(2 \pi)^2}
J (K_{1\perp},K_{2\perp},\tau)~~.
\label{othampl}
\eeqa

There are now two independent patterns of color flow, corresponding to
singlet and octet exchange between the pair and the target. Since we
are interested in the emergence of rapidity gaps, we project the
amplitudes in Eqs.~(\ref{a11}) and (\ref{othampl}) onto singlet exchange by
summing over the colors of the target quark. This is crucial to the present
argument since only upon performing this projection do the four diagrams
acquire the same weights (as in QED), and the
eikonal factors in impact parameter space are similarly reconstructed. Upon
summation over target color the color factors of all diagrams in fact reduce
to that of the first one,
\beq
(T^aT^b)_{AB}(T_bT_a)_{CD}~\delta^{CD} = \half C_F \delta_{AB}
\label{colfac}
\eeq
where $C_F=(N^2-1)/2N$.

The transverse
momentum integrals associated with  $A_2^{(1,2)}$ and with $A_2^{(2,1)}$ in
\eq{othampl} are more complicated than the others, since one more  propagator
depends on the loop momentum. This difficulty is however bypassed  by once
again turning to coordinate space, as in \eq{toimpa}. After the target color
summation of \eq{colfac} we find that the complete two--gluon
color--singlet exchange amplitude,
\beq
{A}_2 (\bfr_\perp,\bfk_\perp) = {\rm i} g^4~ 2 m_t \half C_F
\delta_{AB}~W_2(\bfr_\perp,\bfk_\perp)~V^{\lambda_1,
\lambda_2}_\lambda (z, \bfr_\perp)~~, \label{simprel2}
\eeq
is, just like the single--gluon amplitude of \eq{simprel1}, proportional to the
photon wave function $V^{\lambda_1, \lambda_2}_\lambda (z, \bfr_\perp)$
of \eq{bampl}.  The two--gluon eikonal
factor is
\beq
W_2(\bfr_\perp,\bfk_\perp) = \int \frac{d^2 \bfk_{1\perp}}{(2
\pi)^2} J(K_{1\perp},K_{2\perp},\tau)
\left(1 - {\rm e}^{{\rm i} \bfr_\perp \cdot \bfk_{1\perp}} \right)
\left(1 - {\rm e}^{{\rm i} \bfr_\perp \cdot \bfk_{2\perp}} \right)~~,
\label{eik2tau}
\eeq
where $\bfk_2 = \bfk - \bfk_1$.
The $\bfk_{1\perp}$ integral is now finite, due to the extra
suppression for small $K_{1\perp}$ and $K_{2\perp}$ provided by the eikonal
numerators.  It is thus possible to let the cutoff $\tau \to \infty$, and use
for $J$ the simple expression given in \eq{simpJ}. As a result we see that
the two--gluon eikonal factor is a convolution of two one--gluon factors,
\eq{eik1}, and can be written as
\beqa
W_2(\bfr_\perp,\bfk_\perp) & = & \frac{{\rm i}}{2} \int
\frac{d^2 \bfk_{1\perp}}{(2 \pi)^2} \frac{1 - {\rm e}^{{\rm i}
\bfr_\perp \cdot \bfk_{1\perp}}}{K_{1\perp}^2}
\int \frac{d^2 \bfk_{2\perp}}{(2 \pi)^2} \frac{1 - {\rm e}^{{\rm i}
\bfr_\perp \cdot \bfk_{2\perp}}}{K_{2\perp}^2} \nonumber \\
& \times &
(2 \pi)^2 \delta^2 (\bfk_\perp - \bfk_{1\perp} - \bfk_{2\perp})~~.
\label{eik2}
\eeqa

Transforming $W_2(\bfr_\perp,\bfk_\perp)$ to $\bfR_\perp$-space as in
\eq{besseleik} gives
\beqa
{W}_2(\bfr_\perp,\bfR_\perp) & = & \frac{i}{2} \left[
{W}_1(\bfr_\perp,\bfR_\perp) \right]^2 \nonumber \\
 & = & \frac{i}{8\pi^2} \left[\log\left(\frac{|\bfR_\perp+\bfr_\perp|}
{|\bfR_\perp|} \right) \right]^2~~. \label{logeik}
\eeqa

The fact that the two--gluon eikonal factor $W_2$ is the square of the single
gluon factor $W_1$ is a consequence of the conservation of the impact
parameters $\bfr_\perp,\bfR_\perp$ during the scattering,
and will clearly generalize to an arbitrary number of exchanges
in the singlet channel.
It has an important consequence for the
energy dependence of the cross section, which was first noticed in QED by
Bethe and Maximon \cite{bethemax}. The cross section for single--gluon
exchange involves an integral which is logarithmically divergent at
large impact parameters $\bfR_\perp$ to the target,
\beq
\int d^2 \bfR_\perp |W_1|^2 \propto \int^{2\nu/Q^2} \frac{d^2
\bfR_\perp}{\bfR_\perp^2}~~~.  \label{periint}
\eeq
This results in a cross section which grows logarithmically with the
projectile energy.
For two or more gluon exchanges the integral is on the other hand
convergent at large $\bfR_\perp$, and no logarithm is generated.

\section{Cross--sections}
\label{sigma}

Our results for the one and two--gluon exchange amplitudes $\gamma^*t \to q\bar
q t$ (\ref{comptamp}) can be summarized (see Eqs. (\ref{simprel1})
and (\ref{simprel2})) as
\beq
{A}_n(\bfr_\perp,\bfR_\perp) =
c_n~m_t~W_n(\bfr_\perp,\bfR_\perp)~V(z,\bfr_\perp)~~,
\label{ampn}
\eeq
where $\bfr_\perp,~\bfR_\perp$ are the transverse distances between the
produced quarks and between them and the target, respectively. The photon
wave function $V(z,\bfr_\perp)$ is given in \eq{bampl} and the eikonal
factors $W_n(\bfr_\perp,\bfR_\perp)$ for $n=1,~2$ gluon exchange in Eqs.
(\ref{besseleik}) and (\ref{logeik}), respectively. The coefficients $c_n$,
\beqa
c_1 & = & 2 i g^2 (T^a)_{AB} (T_a)_{CD}~~,\nonumber \\
c_2 & = & i g^4 C_F \delta_{AB}~~,
\label{coeffs}
\eeqa
reflect the color structure.

The spin--averaged square of the full $et \to eq\bar q t$ amplitude $T_n$ can
be expanded as in \eq{atsq},
\beq
\overline{|T_n|^2} = \frac{4 e^4 e_q^2}{\pi^2} \frac{\nu^2 m_t^2}{y^2 Q^2}
\left|c_nW_n(\bfr_\perp,\bfR_\perp)\right|^2 z(1-z) F(z,\bfr_\perp)
\label{tsq}~~,
\eeq
with
\beqa
F(z,\bfr_\perp) & = & \mbox{$\half$} [1+(1-y)^2] \left\{ [1-2z(1-z)]
\varepsilon^2 K_1^2 (\varepsilon r_\perp) +
m^2 K_0^2(\varepsilon r_\perp) \right\}
\nonumber \\
&+& 4(1-y)[z(1-z)]^2Q^2 K_0^2(\varepsilon r_\perp) \nonumber \\
&-& 2(1-y) \cos(2\varphi) z(1-z) \varepsilon^2 K_1^2(\varepsilon r_\perp)~~,
\label{ffac}
\eeqa
where we have used the explicit expression for the photon wave function
derived in the Appendix (see also Ref.~\cite{BHQ}).
Here $\varphi$ is the azimuthal angle
between the lepton plane and the interquark separation $\bfr_\perp$.

The cross--section is given by
\beq
Q^2 \frac{d \sigma}{dQ^2 dx d\varphi} = \frac{1}{(4\pi)^8}
\frac{4 m_N}{m_t^2\nu} \int \frac{d z}{z(1 - z)} d^2 \bfk_\perp
d^2 \bfp_{1\perp} \, \overline{|T|^2}~~,
\label{cross}
\eeq
where $m_t$ is the target mass.
We may thus define a ``cross--section'' in impact parameter space as
\beq
Q^4 \frac{d\sigma_n}{dQ^2dx d\varphi d^2\bfR_\perp} = \frac{\alpha^2e_q^2}
{(2\pi)^4} \frac{Q^2}{xy^2} \int_0^\half dz \int d^2 \bfr_\perp
\left|c_nW_n(\bfr_\perp,\bfR_\perp)\right|^2 F(z,\bfr_\perp),
\label{crossn}
\eeq
whose integral over $\bfR_\perp$ gives the usual DIS cross section.
In the Bjorken scaling limit (\ref{abjscal}) this cross--section should (at
leading twist) be independent of $Q^2$.

In considering the properties of \eq{crossn} it is
important to distinguish the regions of large and small impact parameters
$R_\perp$, corresponding to small and large momentum transfer to the target,
respectively.
Most of the HERA data on rapidity gaps is selected for
small momentum transfer, with the target (proton) diffracting into a low
mass system \cite{hera}. Large momentum transfers implies particles or jets
with large transverse momentum also at target rapidities.
There is experimental evidence for this type of diffractive events as
well \cite{zeuspt}.

\subsection{Low momentum transfer to the target: $R_\perp\gg 1/m$}
\label{sigmalow}

The Bessel functions $K_{0,1}(\varepsilon r_\perp)$ in \eq{ffac} limit the
size of the quark pair to $r_\perp^2 \lsim {\rm min}\{1/m^2,1/z(1-z)Q^2\}$.
Hence for $R_\perp\gg 1/m$ we have also $R_\perp\gg r_\perp$, and the eikonal
factors (\ref{besseleik}) and (\ref{logeik}), integrated over the azimuthal
angle $\varphi_R$ between $\bfR_\perp$ and $\bfr_\perp$, are approximately
\beq
\int_0^{2\pi}d\varphi_R\, |W_n|^2 = a_n \left(\frac{r_\perp}{R_\perp}
\right)^{2n}~~,  \label{wdip}
\eeq
where $a_1=1/4\pi$ and $a_2=3\pi/(4\pi)^4$. The integrals over
$\bfr_\perp$ in \eq{crossn} can then be done using
\beq
\int_0^\infty dr_\perp^2\, r_\perp^{2n}K_j^2(\varepsilon r_\perp) =
\frac{b_{jn}}{\varepsilon^{2 n + 2}}~~,  \label{rintegr}
\eeq
where the $b_{jn}$ $(j=0,1;\ n=1,2)$ are numerical constants. Altogether the
cross section becomes
\beqa
&&\left.Q^4 \frac{d\sigma_n}{dQ^2dx d\varphi dR_\perp^2}
\,\right|_{R_\perp\gg 1/m}
= \frac{\alpha^2 e_q^2}{4(2\pi)^3} \frac{Q^2}{xy^2} |c_n|^2
\frac{a_n}{R_\perp^{2n}} \nonumber \\
&\times& \int_0^{\half} \frac{dz}{\varepsilon^{2n+2}}
\left\{\mbox{$\half$} [1+(1-y)^2] \left[ (1-2z(1-z)) \varepsilon^2
b_{1n} + m^2 b_{0n} \right] \right. \nonumber \\
&+& \left. 4(1-y)[z(1-z)]^2 Q^2 b_{0n} -
2(1-y) \cos(2\varphi) z(1-z) \varepsilon^2 b_{1n} \right\}~~.
\label{signlow}
\eeqa

There are two regions of the final $z$-integral in \eq{signlow} that should
be considered separately.

\begin{itemize}
\item[(a)] $z = \order{m^2/Q^2},~r_\perp = \order{1/m}$.

This is the ``aligned jet region'' \cite{aljet}, relevant to the lowest order
DIS process $\gamma^* q\to q$.
In this region $\varepsilon$ is finite ($\order{m}$), so the exponential
suppression in the Bessel functions in \eq{rintegr} forces $r_\perp$ to be
of order $1/m$. Thus in the scaling limit one of the quarks has a finite
momentum, $z\nu \simeq m^2/2m_Nx$, and the transverse size of the quark pair
also remains finite.
Only terms in the integrand of \eq{signlow} that are of $\order{z^0}$
contribute. Hence the last two terms, which arise from photons with
$\lambda=0$ and from $\lambda = \pm 1$ interference, respectively, can
be ignored, and the leading twist cross section is isotropic in azimuth.
Since $\varepsilon$ is finite, this region contributes both to
the single and to the double--gluon exchange cross sections $(n=1,2)$.
This is as expected, since attaching several gluons to
the quark pair does not change the scaling behavior of the cross section when
the transverse size of the pair is fixed.
\item[(b)] $z = \order{\mbox{$\half$}},~r_\perp = \order{1/Q}$.

In this region both quarks take a finite fraction of the photon momentum,
so that $\varepsilon = \order{Q}$, which in turn forces
the transverse size of the pair to decrease with $Q$.
The photon fragments thus typically have large transverse momentum. Since
$\varepsilon \propto Q$, the contribution from this region to double--gluon
color--singlet exchange $(n=2)$ is suppressed by a factor $1/Q^2$, \ie, it is
of higher twist. This is due to the compactness of the quark pair, which
suppresses gluon attachment. In the case of single gluon exchange the
suppression is compensated by the larger phase space available in $z$,
compared to case (a). In the infinite momentum frame, the contribution of
region (b) with $n = 1$ corresponds to the process $\gamma^* g\to q\bar q$
in inclusive DIS, which is suppressed only logarithmically by the running
coupling $\alpha_s(Q^2).$
\end{itemize}

We conclude that, for moderate momentum transfer to the target,
the color--singlet $(n=2)$ cross section (\ref{signlow}) gets a scaling
contribution only from the aligned jet region (a). After the final
$z$ integration, we find
\beqa
\left.Q^4 \frac{d\sigma_2}{dQ^2dx d\varphi dR_\perp^2}
\,\right|_{R_\perp\gg 1/m}
&=& \frac{\alpha^2 e_q^2}{4(2\pi)^3} \frac{|c_2|^2}{xy^2}
\nonumber \\
&\times& \frac{a_2}{2 R_\perp^4 m^2} \half [1 + (1 - y)^2] \left(2 b_{12}
+ b_{02} \right)~~.  \label{sig2low}
\eeqa
Notice that, according to \eq{sig2low}, the heavy quark diffractive
cross--section decreases with the quark mass $m$ as $1/m^2$.
This is analogous to the suppression of diffractive
production in the symmetric region (b) above. The size of a heavy quark pair
is small, $r_\perp= \order{1/m}$, suppressing the coupling of multiple
gluons. However, heavy
quarks can be produced at leading power in the quark mass
through higher Fock states of the photon, such as $q\bar q g$. The two--gluon
exchange can then occur off the soft gluon in the Fock state, if it has a
finite momentum and thus also a finite transverse size distribution in the
scaling limit (see Section \ref{four}).

It is also interesting to compare our analysis with the
perturbative QCD calculations\cite{BFGMS}
for diffractive vector meson leptoproduction:
$\gamma^* p \to V p.$  In that case the leading
contribution to the cross section arises from
longitudinally--polarized photons and the symmetric regime  $z =
\order{\mbox{$\half$}},~r_\perp = \order{1/Q}$.  Because of the color
cancellations the exclusive diffractive cross section is suppressed by a
factor $1/Q^4$ relative to the Bjorken--scaling rapidity gap rate.

\subsection{Sum over target momentum transfers}
\label{sigmahigh}

In the remainder of this section we will focus on color--singlet exchange.
Since the eikonal factor $W_2(\bfr_\perp,\bfR_\perp)$ according
to Eq. (\ref{logeik}) depends only on the ratio
$|\bfR_\perp+\bfr_\perp|/|\bfR_\perp|$, an integral over all impact
pararameters $\bfR_\perp$ in the cross section (\ref{crossn}) can be readily
done. By dimensional analysis
\beq
\int d^2 \bfR_\perp \, |W_2(\bfr_\perp,\bfR_\perp)|^2 = d_2 r_\perp^2~~,
\label{wint}
\eeq
where $d_2$ is a calculable numerical constant. When
unconstrained, the typical value of $R_\perp$ in the cross section
(\ref{crossn}) is always of $\order{r_\perp}$. Note that the power
of $r_\perp$ in \eq{wint} differs from that of
\eq{wdip} (valid for $R_\perp \gg r_\perp$).
In terms of the constants $b_{jn}$ of
\eq{rintegr} we thus find the $\bfR_\perp$--integrated cross section to be
\beqa
Q^4 \frac{d\sigma_2}{dQ^2dx d\varphi} &=& \frac{\pi\alpha^2
e_q^2}{(2\pi)^4} \frac{Q^2}{xy^2} |c_2|^2 d_2 \int_0^{\half}
\frac{dz}{\varepsilon^4} \nonumber \\
&\times& \left\{\mbox{$\half$} [1+(1-y)^2] \left[ (1-2z(1-z)) \varepsilon^2
b_{11} + m^2 b_{01} \right] \right. \label{signhigh} \\
&+& \left. 4(1-y)[z(1-z)]^2 Q^2 b_{01} -
2(1-y) \cos(2\varphi) z(1-z) \varepsilon^2 b_{11} \right\}~~,
\nonumber
\eeqa

In region (a) of the $z$ integral in \eq{signhigh} ($z=\order{m^2/Q^2}$,
\cf\ Section \ref{sigmalow}), $r_\perp$ (and hence also $R_\perp$) is finite
in the scaling limit. The scattering then dominantly proceeds via moderate
momentum transfer $\order{1/R_\perp}$ to the target. On the other hand,
for finite values of $z$ in \eq{signhigh} the typical values of $R_\perp$ are
of $\order{1/Q}$, implying large momentum transfers to the target. The
corresponding inclusive DIS process is then $\gamma^*t \to q\bar q t$,
where for a realistic target $t$ would be a light quark (or gluon)
constituent. Now the two--gluon exchange process is not suppressed since the
typical hardness of the gluons is commensurate with the size of the $q\bar
q$ dipole. This hard diffractive process (like its inclusive equivalent) is
only suppressed by powers of the running coupling $\as(Q^2)$.

The $z$ integration in \eq{signhigh} can readily be performed, and the
leading twist contribution extracted. We find
\beqa
Q^4 \frac{d \sigma_2}{d Q^2 d x d \varphi} &=& \frac{\pi \alpha^2
e_q^2}{(2\pi)^4} \frac{|c_2|^2 d_2}{xy^2}
\Bigg\{ \frac{1}{2}[1 + (1 - y)^2] \left[b_{01} - b_{11}
\left(1 - \log \frac{Q^2}{m^2} \right) \right]  \nonumber \\
& + & (1 - y) \left( 2 b_{01} - b_{11} \cos(2 \varphi) \right)
\Bigg\}~~.
\label{sigtot}
\eeqa
Notice that for hard diffractive events the azimuthal distribution is not
isotropic, and a logarithmic dependence on $Q^2$ is generated, since
a much larger range in $p_\perp$ is now available.

\subsection{Discussion}

To summarize, there are two quite distinct kinematical regions contributing
to the leading--twist color--singlet exchange cross section, \eq{sigtot}.
For small momentum transfers to the target, or equivalently for large impact
parameter $R_\perp$, our calculation provides a concrete realization of the
aligned jet model \cite{aljet}: asymmetric quark pairs with finite transverse
size dominate in the scaling limit, with the scattering occurring off the
slow quark. In this case the cross section is purely transverse and
independent of azimuth.
If, however, we allow for large momentum trasfers to the target, we
find that the color--singlet cross section receives leading twist contributions
from symmetric and narrow pairs. These contributions have a distinct
azimuthal dependence and rise logarithmically with $Q^2$. They will appear as
events with particles (or jets) of large transverse momentum in both
the current and target fragmentation regions, which balance each other and
are separated by a large rapidity gap. There is already experimental
evidence for such events from ZEUS \cite{zeuspt}.

Our diffractive cross section (\ref{sigtot}) has no contribution
corresponding to events with large transverse momentum jets only in
the current fragmentation region, with low momentum transfer to the target.
Such contributions should occur at higher orders in QCD, for $q\bar qg$
Fock states of the photon (see Section 4).

As we have seen, the low--$x$ cross section is particularly simple when
expressed in impact parameter space. Impact parameters are not directly
measurable, since events are observed in momentum space. For integrated cross
sections this is not an issue, as an integral over all impact parameters is
equivalent to an integral over all transverse momenta. On the other hand,
more differential predictions concerning, \eg, the dependence of the cross
section on the mass $M^2$ of the diffractive  system, and on the size of the
rapidity gap $\Delta \eta$, are less straightforward. At this stage we only
make the following qualitative remarks.

The dependence of the diffractive cross section (\ref{crossn}) on the mass
$M$ of the $(q\bar q)$ system is usually parametrized through
the variable\footnote{In the framework of Pomeron exchange models, $\beta$
is the fraction of Pomeron momentum carried by the struck quark.}
\beq
\beta \equiv \frac{Q^2}{Q^2+M^2} = \frac{\varepsilon^2 - m^2}{\varepsilon^2 +
|\bfp_{1\perp} - z \bfk_\perp|^2}~~,
\label{bdef}
\eeq
where $\varepsilon^2 = m^2+z(1-z)Q^2$, as before. The rapidity gap $\eta$
extends from the target (which is at rest and thus has zero rapidity) to the
slow quark, and (for $z<1/2$) is given by
\beq
\Delta \eta = \log\left(\frac{2z\nu}{m_{1\perp}}\right)~~,
\label{rapg}
\eeq
where $m_{1\perp}=\sqrt{p_{1\perp}^2+m^2}$.
Notice that we use a frame where the $\gamma^*$ is moving along the $z$-axis.

We can estimate the average size of the rapidity gap
in the various kinematical regions contributing to the cross section from the
uncertainty relation corresponding to \eq{toimpa},
\beq
p_{1\perp} \simeq \frac{1}{r_\perp}~~.
\label{uncer}
\eeq
In the aligned jet region we have (as discussed above)
$\langle r_\perp \rangle \sim 1/m$, and hence
$\langle p_{1\perp} \rangle \sim m$.
Since in this region $z \sim m^2/Q^2$ the typical rapidity gap will be
\beq
\Delta \eta \sim \log \left(\frac{m}{m_N x} \right)~~.
\label{tipgap1}
\eeq
For large momentum transfer to the target, on the other hand, region (b)
contributes to the cross section, so that $\langle r_\perp \rangle \sim 1/Q$,
and  consequently $\langle p_{1,\perp} \rangle \sim Q$. Furthermore,
in this region $z \sim 1/2$, so that the size of the gap will now depend
on $Q$. We get
\beq
\Delta \eta \sim \log \left(\frac{Q}{m_N x} \right)~~.
\label{tipgap2}
\eeq
The typical value of $\beta$ remains finite in both regions, as it should.
However we expect region (b) to give a relatively larger contribution at small
values  of $\beta$, since the diffractive system contains large $p_\perp$
jets. Furthermore it should be kept in mind that higher order
contributions from gluon radiation will tend to increase the cross section
at low $\beta$.

The estimates (\ref{tipgap1}), (\ref{tipgap2}) for the size of the rapidity
gap are at the parton level. Hadronization may to some extent fill in the
gap. In the next section we shall address this question by studying the
angular distribution of soft gluon radiation.

\section{Soft Gluon Radiation}
\label{four}

In the case of color--singlet exchange one expects that hadrons will be
produced mainly in the rapidity interval between the quarks, leaving a gap
between the slower quark and the target (here at zero rapidity). While this
hadronization pattern cannot be demonstrated using perturbative methods, it
may be corroborated by studying the distribution of soft (but perturbative)
gluons \cite{chze}. This is analogous to the successful description
\cite{string} of the ``string effect'' in $e^+e^-$ annihilation.

\begin{figure}[htbp]
\begin{center}
\leavevmode
{\epsfxsize=10truecm \epsfbox{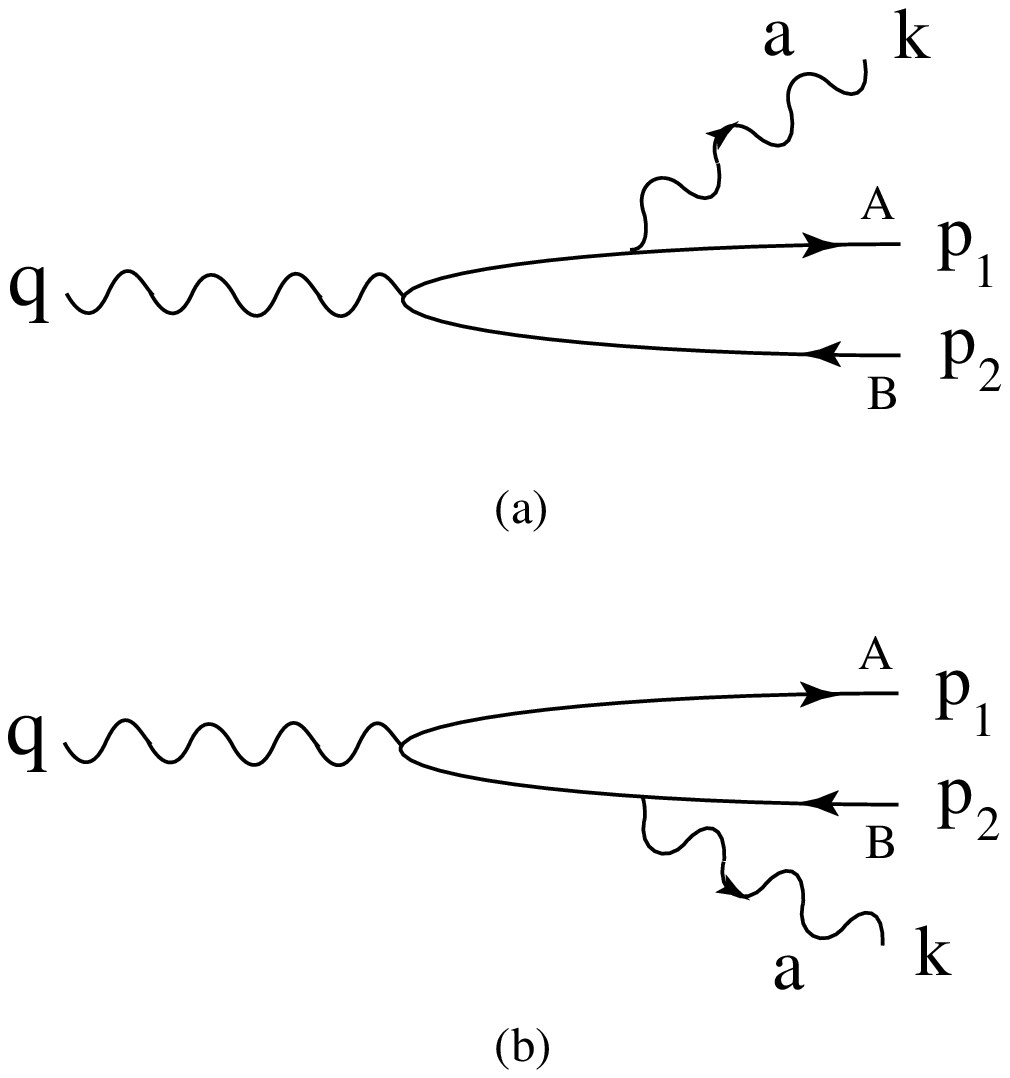}}
\end{center}
\caption[*]{Lowest order diagrams describing the $q\bar q g$ Fock state of
the photon.}
\label{fig4}
\end{figure}

Just as in the $q\bar q$ case above, we expect the gluon distribution in
the final state to be largely determined by its distribution in the photon
wave function. At lowest order, the $q\bar q g$ component of the photon is
given by the two diagrams in Fig. 4. Using the prescription (\ref{arepl})
and the identity (\ref{fident}) to select the dominant time ordering,
we find
\beqa
V(q\bar qg)&=& i\frac{ee_qg T_{AB}^a}{\Delta E} \bar u(p_1)
\left[ \Sslash{\varepsilon}^*(k)\frac{1}{2E_{k1}} \frac{E_{k1}\gamma^0
-(\bfklc+\bfp_1)\cdot \bfgam +m}{E_\gamma-E_2-E_{k1}+i\epsilon}
\bfeps(q) \cdot \bfgam \right. \nonumber \\
&-& \left. \bfeps(q) \cdot \bfgam \frac{1}{2E_{k2}} \frac{E_{k2}\gamma^0
-(\bfklc+\bfp_2)\cdot \bfgam -m}{E_\gamma-E_1-E_{k2}+i\epsilon}
\Sslash{\varepsilon}^*(k) \right] v(p_2) \label{vqqg1}
\eeqa
where $E_\gamma = \sqrt{\bfq^2 - Q^2}$, $E_i = \sqrt{\bfp_i^2 + m^2}$,
$E_{ki} = \sqrt{(\bfklc + \bfp_i)^2 + m^2}$ and $\Delta E = E_\gamma -
E_1 - E_2 - |\bfklc| + {\rm i} \epsilon$.

In the soft gluon limit, $k_z \ll p_{1z},~p_{2z}$ and $|\bfklc_\perp| \ll
|\bfp_{1\perp}|,~|\bfp_{2\perp}|$, the $q\bar q g$ Fock state amplitude
(\ref{vqqg1}) factorizes into a product of the $q\bar q$ amplitude and a
gluon emission factor,
\beq
V(q\bar qg) = V(q\bar q)\, g T_{AB}^a \, \frac{1}{\Delta E} \,
\varepsilon^*(k) \cdot \left(\frac{p_1}{E_1}-\frac{p_2}{E_2} \right)
\label{vqqg2}
\eeq
with $V(q\bar q)$ given by \eq{awfscal}.
The gluon angular distribution is thus governed by the factor
\beq
P(k) = \frac{1}{(\Delta E)^2} \sum_{\lambda_g} \left|
\varepsilon^*_{\lambda_g} (k) \cdot \left(\frac{p_1}{E_1}-\frac{p_2}{E_2}
\right) \right|^2~~,
\label{angfac}
\eeq
where the sum is over the two transverse polarizations available to the gluon.

To evaluate \eq{angfac}, let us for clarity choose the frame so that the gluon
is emitted in, say, the $x-z$ plane, \ie,
$k^\mu = (k, k \sin \theta, 0,
k \cos \theta)$. Within the approximations discussed in Section 2, we find
that the sum over polarizations yields
\beq
\sum_{\lambda_g} \left|\varepsilon^*_{\lambda_g} (k) \cdot
\left(\frac{p_1}{E_1}-\frac{p_2}{E_2} \right) \right|^2 =
\frac{\cos^2 \theta (p_\perp^x)^2 + (p_\perp^y)^2}{\nu^2 z^2 (1 - z)^2}~~,
\label{sumpol}
\eeq
while the energy denominator is given by
\beq
- \Delta E = m_N x + \frac{p_\perp^2 + m^2}{2 \nu z (1 - z)} +
k (1 - \cos \theta)~~.
\label{deleg}
\eeq

For a given $q\bar q$ configuration $(p_\perp,z)$, and a given (soft) gluon
energy $k$, the angular distribution of the gluon will be given by
\beq
P(k) \propto \frac{\cos^2 \theta (p_\perp^x)^2 + (p_\perp^y)^2}{
\left[1 + A (1 - \cos \theta)\right]^2}~~,
\label{angdistr}
\eeq
where
\beq
A = \frac{2 \nu z (1 - z)}{p_\perp^2 + \varepsilon^2} k~~.
\label{BC}
\eeq
In terms of the gluon rapidity
\beq
\eta_g = \frac{1}{2} \log \frac{1 + \cos \theta}{1 - \cos \theta} =
- \log \tan \frac{\theta}{2}~~,
\label{glurap}
\eeq
and of the (slow) quark rapidity $\eta_q$ given by \eq{rapg} we have
\beq
1 + A(1 - \cos \theta) \simeq 1 + \frac{2 k}{m_\perp}
\frac{{\rm e}^{\eta_q}}{1 + {\rm e}^{2 \eta_g}}~~.
\label{rapdistr}
\eeq
This denominator will suppress the probability (\ref{angdistr}) of gluon
emission unless the gluon rapidity is sufficiently high,
\beq
2 \eta_g \gsim \log \left(\frac{2 k}{m_\perp}\right) +\eta_q
\eeq
The softest gluons with $k\simeq m_\perp$ thus have rapidities $\eta_g \gsim
\eta_q/2$. Harder gluons with $k\simeq z\nu$ will be emitted at rapidities
between the fast and the slow quark, $\eta_g \gsim \eta_q$, outside the
rapidity gap to the target.

Thus the gluons associated with the photon Fock state tend to have 
rapidities in the photon fragmentation region, since this minimizes 
the off-shellness of the photon wavefunction.

The above argument concerns gluons that are ``preformed'' in the photon wave
function before the target interaction. The two (Coulomb) gluon interactions
with the target occur during a very short time interval from the point of
view of the photon. Hence, although the projectile system is momentarily a
color octet after the first gluon exchange, there is no formation time
available for (fast) gluon emission before the second gluon is exchanged.
Thus we do not expect that gluons emitted in the short interaction interval
can fill more than a limited part of the rapidity gap, close to target
rapidities.

We have in this paper neglected higher order contributions where the
diffractive scattering occurs off a slow gluon, rather than a quark, in the
photon wave function. The amplitude will still factorize as in \eq{ampn}
into a Fock state amplitude $V$ and a gluon diffractive scattering amplitude
$W$. At lowest order the gluon Fock state amplitude should approximately be
given by  $V(q\bar qg)$ of \eq{vqqg2}. The quark configuration will now be a
typical one, with $z=\order{1/2}$ and $p_\perp^2=\order{Q^2}$. The gluon must
in this case contribute significantly to $\Delta E$ of \eq{deleg}, having a
finite $k_\perp$ in the scaling limit to maintain a constant diffractive
(two--gluon exchange) cross section. Just as in our analysis of quark
diffractive scattering, this implies $k_z = \order{k_\perp^2/m_Nx}$.
The probability for such gluon Fock states scales like $1/Q^2$, as can be
seen by using Eqs.~\ref{sumpol} and \ref{deleg} in the small $\theta$ limit,
\beq
\alpha_s \frac{p_\perp^2}{\nu^2z^2(1-z)^2} \int_0^{\Lambda^2} dk_\perp^2
\int_0^{k_\perp^2/m_Nx} \frac{dk_z}{k_z}
\frac{1}{(m_Nx+k_\perp^2/2k_z)^2} = \order{\alpha_s
\frac{\Lambda^2}{Q^2}}~~.   \label{gprob}
\eeq
Since the gluon diffractive cross section will be independent of $Q^2$ due
to the finite transverse size of the gluon distribution, we have a
scaling contribution to hard diffractive scattering. By including Fock
states  with multiple gluons we may expect to gradually build up the gluon
ladder or ``hard Pomeron'' \cite{bfkl}.

\section{Conclusions}
\label{conc}

In this work we have shown a way to systematically analyze, fully within
perturbative QCD, the physical origin and dynamical dependence of rapidity
gaps in deep inelastic lepton scattering.  Our major simplifying
assumption was the use of a heavy quark target so that the physics of the
PQCD Pomeron could be directly identified with Coulomb gluon exchange between
the heavy target and the constituents of the virtual photon.  Color-octet
exchange from one or more Coulomb gluons coupled to the target is expected to
generate a radiative pattern in the final state  which occupies the entire
rapidity interval between the target and virtual photon fragmentation
regions.  On the other hand, color--singlet exchange arising from two or more
Coulomb gluons corresponds to an absence of soft radiation in the central
rapidity region.

Our calculation shows explicitly the conservation of impact parameters in
DIS at small $x$ \cite{nz}. This leads to a very simple factorized
structure of the scattering amplitude, with the two--gluon eikonal factor
$W_2$ (\eq{logeik}) given by the square of the single gluon factor $W_1$
(\eq{besseleik}).

Due to our use of perturbation theory, the explicit expressions we obtained
for the scattering amplitudes incorporate general features required by
field theory. It is, for example, a well--known feature of QED that single
(Coulomb) photon exchange in Bethe--Heitler lepton pair photoproduction gives
a cross section that increases logarithmically with photon energy $\nu$,
\beq
\sigma_{\gamma Z \rightarrow
\ell^+\ell^- Z} \sim \frac{\alpha(Z\alpha)^2}{m^2_\ell}\ \ell n\,
\frac{\nu}{m_\ell}~~.  \label{qedlog}
\eeq
The logarithm is due to an integration over large impact parameters
$R_\perp$, the upper limit being kinematically given by $R_\perp <
1/K_z^{\rm min} = m^2_\ell/2\nu$. In our approach this logarithm is
a direct consequence of the fact that $W_1 \propto r_\perp/R_\perp$ for
large target impact parameters $R_\perp$, \cf\ \eq{periint}. The fact that
multiple Coulomb photon exchange does not give logarithmically
enhanced contributions \cite{bethemax} then follows directly from \eq{logeik},
$W_2 \propto W_1^2 \propto r_\perp^2/R_\perp^2$, and its generalizations.
This difference in impact parameter dependence shows that the inclusive
and diffractive DIS processes are dynamically distinct, and that their ratio
cannot be characterized by a single color--dependent constant.

In QED the logarithm of \eq{qedlog} saturates at $R_\perp = \order{1\
{\rm \AA}}$, where the scattering becomes coherent over the entire
electrically neutral atom. In QCD the saturation for gluon exchange will
occur at the color confinement radius $R_\perp = \order{1\ {\rm fm}}$, where
perturbative methods cease to apply.

In analyzing the structure of our diffractive cross section (\ref{crossn})
we found it important to distinguish between the case of fixed (small)
momentum transfer $\bfk$ to the target, and scattering involving a target
momentum transfer that increases with the virtuality $Q^2$ of the photon.

For small momentum transfers $\bfk$ we found that only the ``aligned jet''
region contributes to diffractive scattering in the scaling limit.
This region is characterized by one of the quarks taking nearly
all of the photon momentum, and by the transverse size $r_\perp$
of the quark pair remaining finite in the scaling limit.
Hence multiple gluons can couple to the pair at leading twist. This
is in contrast to the ``symmetric jet'' region where the vanishing size of the
pair allows only a single gluon to couple. Since the size of a heavy quark
pair is $r_\perp \lsim \order{1/m}$, two--gluon exhange is similarly
suppressed in heavy quark production, leading to a diffractive
cross section proportional to $1/m^2$.
Analogously, the production of large $p_\perp$ jets at the photon vertex is
suppressed by $1/p_\perp^2$ in diffractive scattering. It will be important
to search for these effects experimentally. These features are, however, true
only at lowest order, when transverse  gluon radiation is neglected.

Most of the HERA data \cite{hera} has been selected for low--mass target
diffraction, which kinematically allows for a long rapidity gap
between the photon and target fragments. It is clearly of interest to study
also whether the hardness of the virtual photon can be transmitted to the
target fragments in diffractive events \cite{zeuspt}.
This would not be expected in a model where the scattering occurs via
``soft'' Pomeron exchange, similar to
that observed in hadron scattering at low momentum transfers. In our
approach we found, to the contrary, that the contribution from small target
impact parameters $R_\perp$ is quite important and only suppressed by
powers of the running coupling $\as(Q^2)$. In inclusive DIS the
corresponding subprocess would be classified as $\gamma^*t \to q\bar q t$,
where $t$ can be any parton (light quark or gluon) of the target. Once
$R_\perp = \order{r_\perp}$, the exchange of multiple gluons to the $q\bar
q$ pair is not power suppressed no matter how small $r_\perp$ is, and the
diffractive process can proceed at leading twist. In this large momentum
transfer region perturbative calculations are actually the most
reliable.

We also considered the modifications brought to the above
picture by radiative gluons. Soft gluons emitted in
color--singlet exchange processes will mostly occupy the rapidity
region between the quarks produced at the photon vertex, thus leaving intact
a central rapidity gap extending to the target fragmentation region. At
higher orders there are also contributions from photon Fock states where the
quark pair is in a transversally compact ``symmetric jet'' configuration while
a soft gluon maintains a finite transverse distance to the quark pair in the
scaling limit \cite{class}. The soft gluon can scatter diffractively (through
two--gluon exchange), allowing heavy quarks and large
$p_\perp$ jets to be produced at the photon vertex
at leading twist. These contributions tend to
have large diffractive mass, and thus contribute to the region of low
$\beta=Q^2/(Q^2+M^2)$. They are, in fact, the beginning of the gluon ladder
which presumably generates the BFKL Pomeron \cite{bfkl}. It should be
possible to extend the present calculation in this direction -- the more
rungs in the ladder that are considered the less important should be our
assumption of a heavy quark target.

\bigskip\noindent
{\bf Acknowledgements}

We are grateful for useful discussions with W. Buchm\"uller and A. Hebecker.

\section*{Appendix}

Here we wish to collect some useful formulas concerning electroproduction
and the virtual photon wave function.

\bigskip
\begin{figure}[htbp]
\begin{center}
\leavevmode
{\epsfxsize=10truecm \epsfbox{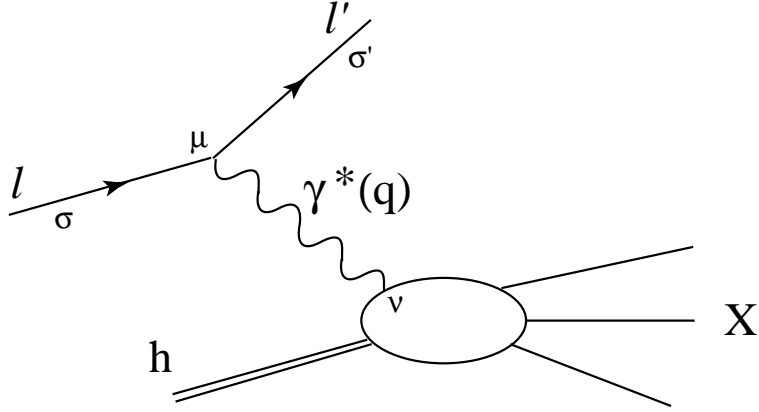}}
\end{center}
\caption[*]{The amplitude for deep inelastic lepton scattering, $\ell+h \to
\ell' + X$.}
\label{figA1}
\end{figure}

\vskip .5cm

\noindent{\it {\bf A1. Virtual Photon Factorization.}} \vskip .3cm

In the notation of Fig. A1, the lepton scattering amplitude is
\beq
T(\ell h \to \ell^\prime h^\prime) = \frac{1}{Q^2}~j_\ell^\mu(-g_{\mu\nu})
j_h^\nu~~,
\label{aampl}
\eeq
where $j_\ell^\mu = e~\bar u_{\sigma^\prime}(\ell^\prime)\gamma^\mu
u_\sigma(\ell)$ is the lepton current and $j_h^\nu$ is the target (hadron)
vertex. We choose the target $(h)$ rest frame where the virtual photon
momentum is aligned with the $z$-axis,
\beq
q = (\nu,{\bf 0}_\perp,\sqrt{\nu^2+Q^2})~~.
\label{aqvec}
\eeq
Gauge invariance, $q^\mu j_\mu =0$, implies
\beq
j_{\ell,h}^0 = \frac{q^z}{q^0} j_{\ell,h}^z~~.
\label{acurr}
\eeq
Eliminating the $\mu,\ \nu =0$ components in \eq{aampl} we can express the
photon exchange in terms of its transverse and longitudinal polarizations,
\beq
T=\frac{1}{Q^2}\sum_{\lambda=\pm 1,0} (-1)^{\lambda+1} \bfj_\ell \cdot
\bfeps^*(\lambda)\ \bfeps(\lambda) \cdot \bfj_h~~,
\label{apol}
\eeq
where
\beqa
\bfeps(\lambda=\pm 1) & = & -\frac{1}{\sqrt 2}(1,\pm i,0)~~,
\nonumber \\
\bfeps(\lambda=0) & = & \frac{Q}{\nu}(0,0,1)~~.
\label{apolvec}
\eeqa
We define the azimuthal angular orientation by fixing the hadronic vertex
at $\varphi=0$, while
\beq
\ell^\prime = (\nu(1-y)/y,\ell_\perp \cos\varphi,\ell_\perp
\sin\varphi,\ell^{\prime z})~~,
\label{aellp}
\eeq
where $\nu=y \ell^0$.

The following expressions will be given in the scaling limit,
\beq
\left\{ \begin{array}{ll}
              Q^2 \to \infty \\
              \nu \to \infty
              \end{array} \right.
\ \ \ {\rm with}\ \ \ x=\frac{Q^2}{2m_N \nu}\ \ \ {\rm fixed},
\label{abjscal}
\eeq
where $m_N$ is the nucleon mass. We have then
\beqa
\ell_\perp &=& \ell_\perp^\prime = Q \sqrt{1-y}/y~~, \\ \nonumber
\ell^{\prime z} &=& \nu (1-y)/y~~.
\label{aellexpr}
\eeqa

The spin--averaged square of the scattering amplitude $T$ in \eq{apol} can be
written
\beq
\overline{|T|^2} = \frac{1}{Q^4} \sum_{\lambda, \lambda'}
(-1)^{\lambda+\lambda'} L_{\lambda\lambda'} H_{\lambda\lambda'}~~,
\label{atsq}
\eeq
where
\beq
L_{\lambda\lambda'} = \frac{1}{2} \sum_{\rm spins} \bfj_\ell \cdot
\bfeps^*(\lambda) \ \bfj_\ell^* \cdot
\bfeps(\lambda') = L_{\lambda'\lambda}^*~~.
\label{allp}
\eeq
Explicitly,
\beqa
L_{\pm1,\pm1} &=& e^2Q^2 \left[1+(1-y)^2 \right] /y^2~~,  \\ \nonumber
L_{\pm1,\mp1} &=& 2e^2Q^2 e^{\mp 2i\varphi} (1-y)/y^2~~,  \\ \nonumber
L_{\pm1,0} &=& -\sqrt{2}e^2Q^2 e^{\mp i\varphi} \sqrt{1-y}(2-y)/y^2~~,  \\
\nonumber
L_{0,0} &=& 4e^2Q^2 (1-y)/y^2~~.
\label{alexpl}
\eeqa
The hadronic tensor is
\beq
H_{\lambda'\lambda} = \frac{1}{2} \sum_{\rm spins} \bfj_h^* \cdot
\bfeps^*(\lambda') \ \bfj_h \cdot
\bfeps(\lambda) = H_{\lambda\lambda'}^* = H_{-\lambda',-\lambda}~~,
\label{ahhp}
\eeq
where the last equality applies if we choose $H$ to be real, which is
always possible by a choice of frame.
With the expressions (\ref{alexpl}) for $L_{\lambda\lambda'}$ we find for
$\overline{|T|^2}$ in \eq{atsq},
\beqa
\overline{|T|^2} &=& \frac{4e^2}{y^2Q^2} \left\{\frac{1}{2} [1+(1-y)^2]~H_{11}
+(1 - y) H_{00} \right. \\ \nonumber
& + & \left. (2 - y) \sqrt{2 (1 - y)} \cos\varphi~{\rm Re} (H_{10}) +
(1-y)\cos(2\varphi)~H_{1,-1} \right\}~~,
\label{atgen}
\eeqa
which is the general expression for the hadronic tensor assuming only single
photon exchange between the lepton and hadron vertices (see for
example Ref.~\cite{rump}).

\vskip .5cm

\noindent {\it {\bf A2. Virtual Photon Wave Function.}} \vskip .3cm

The standard covariant expression for the $\gamma \to q\bar q$ vertex at
lowest order (Fig. A2) is
\beq
iT_\lambda^{\lambda_1\lambda_2} = \bar u_{\lambda_1}(p_1)
\left[-iee_q \delta_{AB} (-\bfeps(\lambda) \cdot \bfgam)
\right] v_{\lambda_2}(p_2) 2\pi\delta(E_\gamma-E_1-E_2)~~,
\label{agamvert}
\eeq
where $E_\gamma \equiv \nu$ is the (virtual) photon energy,
$E_i=\sqrt{\bfp_i^2 + m^2}$ are the quark energies, $A,B$ the quark colors,
and we suppressed the 3--momentum conserving $\delta$--functions.
The photon polarization vectors
$\bfeps(\lambda)$ are given in by \eq{apolvec}, and include the
longitudinal $(\lambda=0)$ polarization for virtual photons with $q^2 =
-Q^2 \neq 0$.

\bigskip
\begin{figure}[htbp]
\begin{center}
\leavevmode
{\epsfxsize=10truecm \epsfbox{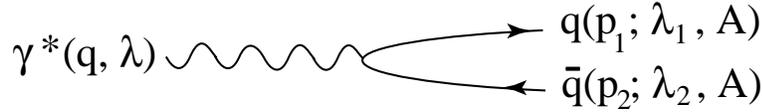}}
\end{center}
\caption[*]{Lowest order diagram describing the $q\bar q$ Fock
state of the photon.}
\label{figA2}
\end{figure}

We shall define the photon wave function $V_\lambda^{\lambda_1\lambda_2}
\delta_{AB}$ by \eq{agamvert} with the replacement
\beq
2\pi\delta(E_\gamma-E_1-E_2) \to \int_{-\infty}^0 dt
\exp[ - {\rm i} t (E_\gamma - E_1 - E_2 + {\rm i} \epsilon)]~~,
\label{arepl}
\eeq
implying
\beq
iV_\lambda^{\lambda_1\lambda_2}\delta_{AB} =
\frac{- e e_q \delta_{AB}}{E_\gamma - E_1 - E_2 + i\epsilon}
\bar u_{\lambda_1}(p_1)
\bfeps(\lambda) \cdot \bfgam\, v_{\lambda_2}(p_2)~~.
\label{awf}
\eeq
Physically, $V_\lambda^{\lambda_1\lambda_2}$ is the lowest order
amplitude for finding a $q\bar q$ pair at a given time $(t=0)$,
given a bare photon at $t=-\infty$. The time $t$ of the transition $\gamma \to
q\bar q$ is summed over the interval $-\infty < t < 0$.

The concept of a photon wave function is natural in the scaling limit
(\ref{abjscal}), where the mass $Q$ of the photon is much less than its
energy $E_\gamma=\nu$. In this limit, \eq{awf} simplifies to
\beq
V_\lambda^{\lambda_1\lambda_2}(z,\bfp_\perp) = -iee_q
\frac{2\nu z(1-z)}{p_\perp^2+\varepsilon^2}
\bar u_{\lambda_1}(p_1) \bfeps(\lambda) \cdot
\bfgam\, v_{\lambda_2}(p_2)~~,
\label{awfscal}
\eeq
where $z$ is the fraction of the longitudinal momentum carried by the
quark,
\beqa
\bfp_1 &=& z\bfq + \bfp_\perp~~, \\ \nonumber
\bfp_2 &=& (1-z)\bfq - \bfp_\perp~~,
\label{azdef}
\eeqa
while
\beq
\varepsilon^2 = m^2 + z(1-z)Q^2~~.
\label{aepsdef}
\eeq
We shall use the helicity basis for the Dirac spinors,
\beqa
u_\lambda(p) &=& \frac{\Sslash{p}+m}{\sqrt{E+m}}
\left( \begin{array}{c} \chi_{_\lambda}(p) \\ 0 \end{array} \right)~~, \\
\nonumber
v_\lambda(p) &=& \frac{-\Sslash{p}+m}{\sqrt{E+m}}
\left( \begin{array}{c} 0 \\ \chi_{_{-\lambda}}(p) \end{array} \right)~~,
\label{ahelsp}
\eeqa
where the 2-component spinors $\chi_{_\lambda}(p)$ are given in terms of the
polar and azimuthal angles $\theta,\ \varphi$ of $\bfp$ as
\beqa
\chi_{_\half}(p) &=&
\left( \begin{array}{c} \cos(\theta/2) \\ e^{i\varphi}\sin(\theta/2)
\end{array} \right)~~, \\ \nonumber
\chi_{_{-\half}}(p) &=&
\left( \begin{array}{c} -e^{-i\varphi}\sin(\theta/2) \\ \cos(\theta/2)
\end{array} \right)~~.
\label{atwosp}
\eeqa

In the scaling limit (\ref{abjscal}) we have
\beqa
& & V_{\lambda=\pm1}^{\lambda_1,\lambda_2}(z,\bfp_\perp)~~=~~
-i e e_q\frac{\nu\sqrt{2z(1-z)}}{p_\perp^2+\varepsilon^2}
\label{atransvert} \\
& \times & \left\{ p_\perp e^{i\lambda\varphi} \delta_{\lambda_1,-\lambda_2}
\left[ z(\lambda-2\lambda_2)+(1-z)(\lambda+2\lambda_2) \right]
-2m \delta_{\lambda,2\lambda_1} \delta_{\lambda_1,\lambda_2} \right\}~~,
\nonumber
\eeqa
and
\beq
V_{\lambda=0}^{\lambda_1,\lambda_2}(z,\bfp_\perp) =
-iee_q\frac{4\nu Q[z(1-z)]^{3/2}}{p_\perp^2+\varepsilon^2}
2\lambda_1 \delta_{\lambda_1,-\lambda_2}~~.
\label{alongvert}
\eeq

It is useful to express the photon wave function in impact parameter
space, through the Fourier transform
\beq
{V}^{\lambda_1, \lambda_2}_\lambda (z, \bfr_\perp) =
\int \frac{d^2 \bfp_\perp}{(2 \pi)^2}~{\rm e}^{{\rm i} \bfr_\perp \cdot
\bfp_\perp}~ V^{\lambda_1, \lambda_2}_\lambda (z, \bfp_\perp)~~.
\label{FT}
\eeq
The integral in \eq{FT} can be explicitly performed, for the photon
wave functions in Eqs.~(\ref{atransvert}) and (\ref{alongvert}), as well as
for  the other amplitudes discussed in the text, by using the identity
\beq
\int_0^\infty d u \frac{u^{n + 1}}{u^2 + a^2} J_n(u) = a^n K_n(a)~~,
\label{besselid}
\eeq
where $J_n$ and $K_n$ are Bessel functions, and the identity applies for
$n < 3/2$. The results can be written as
\beqa
{V}^{\lambda_1, \lambda_2}_{\lambda = \pm 1} (z, \bfr_\perp) & = &
\frac{{\rm -i}}{2 \pi}~e e_q~\nu \sqrt{2 z (1 - z)}
\left\{-2m~\delta_{\lambda,2\lambda_1}~\delta_{\lambda_1,\lambda_2}
K_0(\varepsilon r_\perp) \right. \nonumber \\
 & + & \left. {\rm i~e}^{{\rm i}\lambda \varphi_r}
\delta_{\lambda_1, - \lambda_2}
\left[z (\lambda - 2 \lambda_2) + (1 - z)(\lambda + 2 \lambda_2) \right]
\varepsilon K_1(\varepsilon r_\perp) \right\}
\nonumber \\
{V}^{\lambda_1, \lambda_2}_{\lambda = 0} (z, \bfr_\perp) & = &
\frac{{\rm -i}}{2 \pi}~e e_q~\nu~Q~
4 \left[z (1 - z) \right]^{3/2} 2
\lambda_1~\delta_{\lambda_1, - \lambda_2}~K_0(\varepsilon r_\perp)~~,
\label{bampl}
\eeqa
where $\varphi_r$ is the azimuthal angle of $\bfr_\perp$.

As shown in Section 2, the scattering amplitudes with one and two gluon
exchanges off a massive target are, in the small $x$ limit, proportional
to the photon wave functions in \eq{bampl}.

\end{document}